\documentclass[showkeys,aps,amssymb,prd]{revtex4}
\usepackage{bm,graphicx}

\newcommand\beq{\begin{equation}}
\newcommand\beqa{\begin{eqnarray}}
\newcommand\beqan{\begin{eqnarray*}}
\newcommand\eeq{\end{equation}}
\newcommand\eeqa{\end{eqnarray}}
\newcommand\eeqan{\end{eqnarray*}}

\newcommand\gravr{{\sf m}_\bullet}
\newcommand\bhpm{{\cal M}_\bullet}
\newcommand\bha{{\sf a}}
\newcommand\bhpJ{{\cal J}_\bullet}
\newcommand\cE{{\cal E}}
\newcommand\cL{{\cal L}}

\newcommand\cQ{{\cal Q}}
\newcommand\cur{{\cal C}}

\newcommand\hl{\hat{\lambda}}
\newcommand\hcL{\hat{\cal L}}
\newcommand\hcQ{\hat{\cal Q}}

\newcommand\vth{\vartheta}
\newcommand\vphi{\varphi}

\newcommand\snq{{\tt q}}

\newcommand\cb{{\cal B}}

\newcommand\cd{{\tt d}}

\newcommand\ahat{\hat{a}}
\newcommand\kpolar{\zeta}
\newcommand\kazym{\phi}

\newcommand\reffig[1]{Fig.~\ref{fig:#1}}

\newcommand\refsec[1]{Section~\ref{sec:#1}}

\newcommand\refapp[1]{Appendix~\ref{app:#1}}

\begin{document}

\title{Lensing by Kerr Black Holes. I: General Lens Equation and Magnification Formula}

\author{Amir B.\ Aazami}
\affiliation{Department of Mathematics,\\ Duke University,\\
  Science Drive, Durham, NC 27708-0320;\\
  {\tt aazami@math.duke.edu}}

\author{Charles R.\ Keeton}
\affiliation{Department of Physics \& Astronomy, Rutgers University,
  136 Frelinghuysen Road, Piscataway, NJ 08854-8019;\\
  {\tt keeton@physics.rutgers.edu}}

\author{A.\ O.\ Petters}
\affiliation{Departments of Mathematics and Physics, Duke University,\\
  Science Drive, Durham, NC 27708-0320;\\
  {\tt petters@math.duke.edu}}
  
\begin{abstract}
We develop a unified, analytic framework for 
gravitational lensing by Kerr black holes.  In this first paper we present
a new, general lens equation and magnification formula governing lensing by a compact object.
Our lens equation assumes that the source and observer are in the asymptotically
flat region and does not require a small angle approximation.  Furthermore, it takes into account the displacement that occurs when the light ray's tangent lines
at the source and observer do not meet on the lens plane.  
We then explore our lens equation in the case when the compact object is a Kerr black hole.
Specifically, we give an explicit expression for the
displacement when the observer is in the equatorial plane of the Kerr black hole as well as
for the case of spherical symmetry.
\end{abstract}
\maketitle

\section{Introduction}
\label{sec:intro}

The field of gravitational lensing has seen exponential growth
in its physical \cite{SEF,KSW} and mathematical \cite{PLW,PWreview,Petters} infrastructure,
yielding diverse applications in astronomy and cosmology.
In this paper we address gravitational lensing in the setting of one of the
most important non-spherically symmetric and non-static solutions
of the Einstein equations, namely, Kerr black holes.  This has already been the focus of many studies.  Indeed, several authors have explored the Kerr's caustic structure, as well as Kerr black hole lensing in the
strong deflection limit, focusing on leading order effects in light
passing close to the region of photon capture
(e.g., Rauch \& Blandford \cite{RB}, Bozza \cite{Bozza,Bozza2,Bozza3}, V\'asquez \& Esteban \cite{VqzE}, 
Bozza, De Luca, Scarpetta, and Sereno \cite{Betal}, Bozza, De Luca, and Scarpetta \cite{Betal2}, and Bozza \& Scarpetta \cite{BS2}).

Studies of Kerr lensing have also been undertaken in the weak deflection limit.  In particular, Sereno \& De Luca \cite{SerenoDeLuca,SerenoDeLuca2} gave an analytic treatment of caustics and two lensing observables for Kerr lensing in the weak deflection limit, while Iyer \& Hansen \cite{Iyer1,Iyer2} found an asymptotic expression for the equatorial bending angle.  Werner \& Petters \cite{WernerPetters} used magnification relations for weak-deflection Kerr lensing to address the issue of Cosmic Censorship (for lensing and Cosmic Censorship in the spherically symmetric case, see Virbhadra \& Ellis \cite{VE2}).  

In Papers I and II of our series, we are developing a comprehensive analytic framework for Kerr black
hole lensing, with a focus on regimes
beyond the weak deflection limit (but not restricted to the strong
deflection limit).  In three earlier papers \cite{KP1,KP2,KP3}, Keeton \& Petters
studied lensing by static, spherically symmetric compact objects in
general geometric theories of gravity.  In \cite{KP1,KP2}, the authors found
universal relations among lensing observables for Post-Post-Newtonian (PPN) models that
allowed them to probe the PPN spacetime geometry beyond the
quasi-Newtonian limit.  In \cite{KP3} they considered braneworld gravity,
which is an example of a model outside the standard PPN family as it
has an additional independent parameter arising from an extra
dimension of space.  They developed a wave optics theory (attolensing)
to test braneworld gravity through its signature in interference
patterns that are accessible with the Fermi Gamma-ray Space
Telescope.  Papers I and II present a similar analysis of Kerr black hole lensing beyond the weak deflection limit.

The outline of this paper is as follows.  In Section~\ref{sec:gen-lenseq} we present
a new, general lens equation and magnification formula governing lensing by a thin deflector.
Both equations are applicable for non-equatorial observers and 
assume that the source and observer are in the asymptotically
flat region.
In addition, our lens equation incorporates 
the displacement for a general setting that Bozza \& Sereno \cite{BS} 
introduced for the case of a spherically symmetric deflector.
This occurs when the light ray's tangent lines
at the source and observer do not intersect on the lens plane.
Section~\ref{sec:lenseq-Kerr} gives an explicit expression for this
displacement when the observer is in the equatorial plane
of a Kerr black hole as well as
for the case of spherical symmetry.   The lens equation itself is derived in Appendix~\ref{app:Kerr-null-geodesics}.

In Paper II we solve our lens equation perturbatively to obtain analytic
expressions for five lensing observables (image positions, magnifications,
total unsigned magnification, centroid, and time delay) for the regime of
quasi-equatorial lensing.

\section{General Lens Equation with Displacement}
\label{sec:gen-lenseq}

\subsection{Angular Coordinates on the Observer's Sky}
\label{sec:obs-coords}

Let us define Cartesian coordinates $(x,y,z)$ centered on the
compact object and oriented such that the observer lies on the
positive $x$-axis.  We assume that gravitational lensing will take place outside the photon region, which is outside the ergosphere, so that $(x,y,z)$ are always spatial coordinates.  We also point out that we will not be considering distances more than a Hubble time, so that our formalism ignores the expansion of the universe.

\begin{figure}[t]
\begin{center}
\includegraphics[scale=.64]{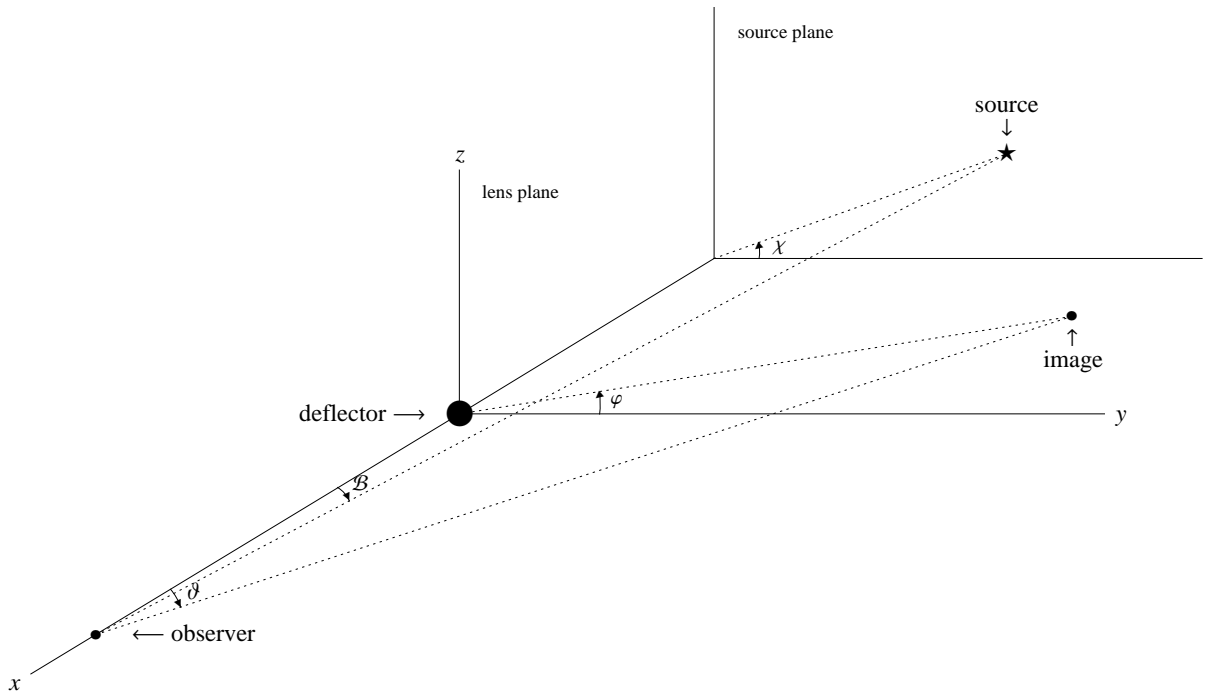}
\end{center}
\caption{
Angles on the observer's sky.
An image's position is determined by $(\vth, \vphi)$.  The source's position is given by $(\cb,\chi)$.}
\label{fig:ObsCoords}
\end{figure}

Assume that the observer in the asymptotically flat region is at
rest relative to the $(x,y,z)$ coordinates.  {\em All equations
derived in this section are relative to the asymptotically flat
geometry of such an observer.}
The natural coordinates for the observer to use in gravitational
lensing are angles on the sky.  To describe these angles, we
introduce ``spherical polar" coordinates
defined with respect to the observer and the optical axis (from the observer to
the lens), and the $yz$-planes at the deflector and the light source.  The vector to the image position is then described by the
angle $\vth$ it makes with the optical axis, and an azimuthal
angle $\vphi$.  Similarly, the vector to the source position is
described by the angle $\cb$ it makes with the optical axis and
by an azimuthal angle $\chi$.  These angles are shown in
\reffig{ObsCoords}.  Note that the optical axis is the $x$-axis.
We adopt the usual convention for spherical polar coordinates:
the image position has $\vth > 0$ and $0 \le \vphi < 2\pi$,
while the source position has $\cb \ge 0$ and $0 \le \chi < 2\pi$.
In fact, since we only need to consider the ``forward'' hemisphere from the observer
we can limit $\vth$ to the interval $(0,\pi/2)$ and $\cb$ to the interval $[0,\pi/2)$.

The
``lens plane'' is the plane perpendicular to the optical axis
containing the lens, and the ``source plane'' is the plane
perpendicular to the optical axis containing the source; these are also shown in \reffig{ObsCoords}.  Define
the distances $d_L$ and $d_S$ to be the perpendicular distances
from the observer to the lens plane and source plane, respectively,
while $d_{LS}$ is the perpendicular distance from the lens plane
to the source plane.  Some investigators define $d_S$ to be the
distance from the observer to the source itself, as opposed to
the shortest distance to the source plane.  We shall comment on
this distinction in Section~\ref{sec:spherical}.

\subsection{General Lens Equation via Asymptotically Flat Geometry}
\label{sec:gen-lens-eqn}

\begin{figure}[t]
\begin{center}
\includegraphics[scale=.67]{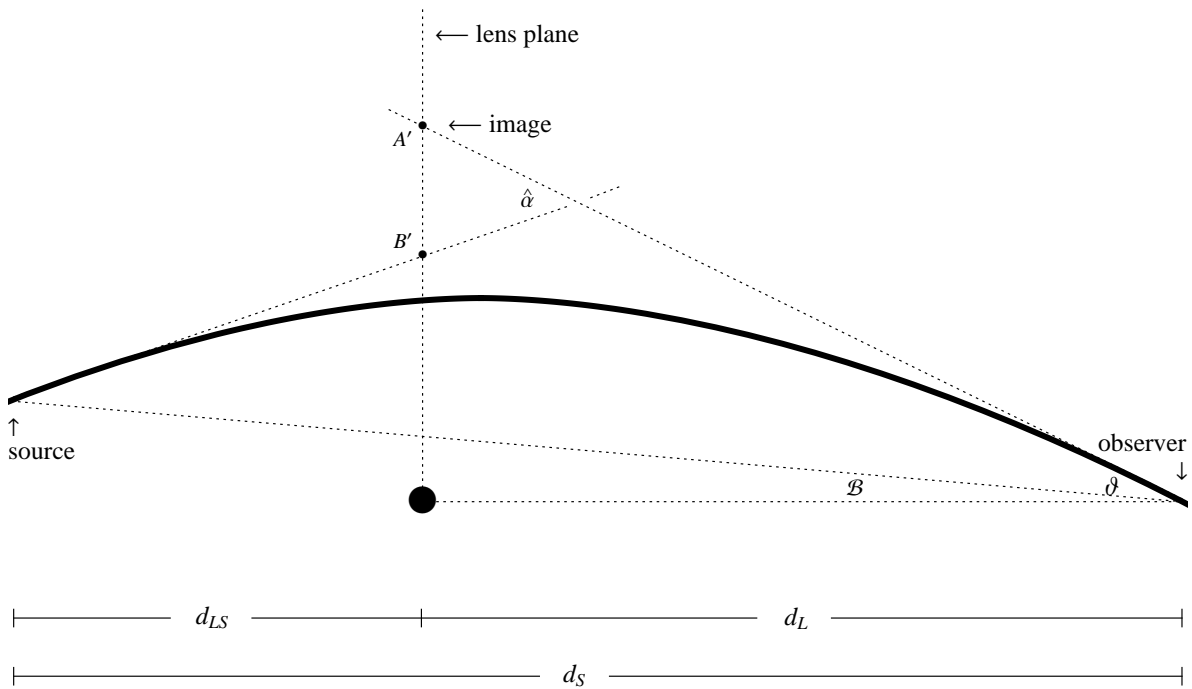}
\end{center}
\caption{
A lensing scenario demonstrating that the tangent line to the segment of the ray arriving at the observer and the tangent line of the ray at the source need not intersect on the lens plane; i.e., $A' \neq B'$ in general.  The angles $\cb$ and $\vartheta$ are as in \reffig{ObsCoords} (or rather, they are their projections onto the $xz$-plane), $\hat{\alpha}$ is the ``bending angle," and $d_L$, $d_{S}$, and $d_{LS}$ are the perpendicular distances between the lens plane and observer, the source plane and observer, and the lens and source planes, respectively.}
\label{fig:LensGeom}
\end{figure}

\begin{figure}[t]
\begin{center}
\includegraphics[scale=.62]{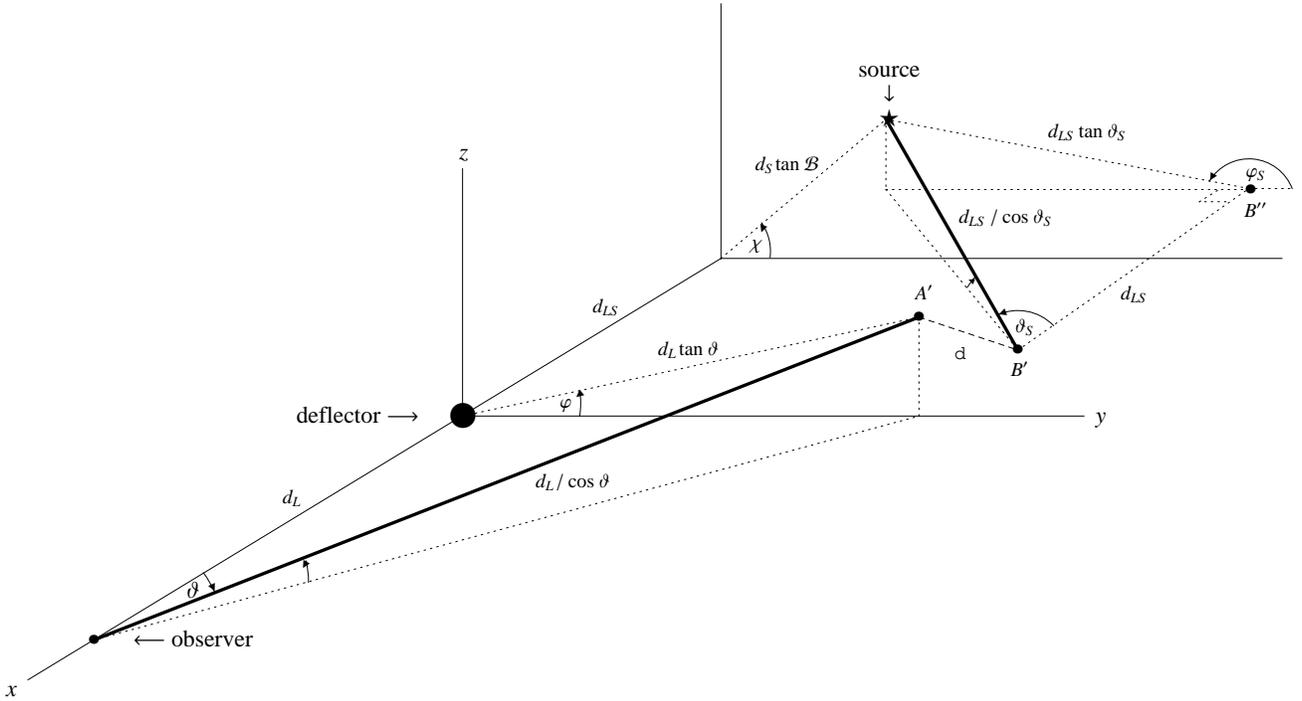}
\end{center}
\caption{
A detailed diagram of lensing with displacement.  The tangent line to the segment of the ray arriving at the observer is determined by $(\vth,\vphi)$ and intersects the lens plane at $A'$, while the tangent line to the ray emanating from the source is determined by $(\vth_S,\vphi_S)$ and intersects the lens plane at $B'$.  The distance between these two points is quantified by the displacement amplitude $\cd$, whose horizontal and vertical components we denote by $\cd_y$ and $\cd_z$, respectively.  The deflector could be a Kerr black hole and the light ray may dip below the $xy$-plane.}
\label{fig:Lens-Geom3}
\end{figure}

Consider the lensing geometry shown in \reffig{LensGeom}.  With respect to the light ray being lensed, there are two tangent lines of interest: the tangent line to the segment of the ray arriving at the observer and the tangent line to the ray emanating from the source.  
As first emphasized in \cite{BS}, {\it it is important to realize that
these two tangent lines need not intersect.}  If they do intersect (as
for a spherical lens, since in that case the tangent lines are
coplanar), the intersection point need not lie in the lens plane.
This effect has often been neglected, and while it may be small in the
weak deflection limit (see Section~\ref{sec:spherical} below) we should include it for greater generality.
A simple way to capture this displacement is to
consider the points where the two tangent lines cross the lens
plane, namely, the points $A'$ and $B'$ in \reffig{LensGeom}.
If the tangent lines do intersect in the lens
plane, then $A' = B'$.  Otherwise, as can be seen in greater detail in \reffig{Lens-Geom3}, there is a displacement on the lens plane that
we quantify by defining
\beq \label{eq:disp-def}
\cd_y = B'_y - A'_y\,, \qquad \cd_z = B'_z - A'_z\,.
\eeq
Note from \reffig{Lens-Geom3} that the
tangent line to the segment of the ray arriving at the observer 
is determined by $(\vth,\vphi)$.  The tangent line to the ray
emanating from the source can likewise be described by the angles
$(\vth_S,\vphi_S)$, where $-\pi/2 < \vth_S < \pi/2$ and $0 \leq \vphi_S < 2\pi$.  As shown in \reffig{Lens-Geom3}, $\vth_S$ has vertex $B'$ and is measured from the line joining the points $B'$ and $B''$, which runs parallel to the optical axis.  We adopt the following sign convention for $\vth_S$: if $\vth_S$ goes {\it toward} the optical axis, then it will be positive; otherwise it is negative (e.g., the $\vth_S$ shown in \reffig{Lens-Geom3} is positive).
We will obtain the general lens equation by considering the coordinates
of the points $A'$ and $B'$ in \reffig{Lens-Geom3}.  Using the
asymptotically flat geometry of the observer, we have
\beqa
A'_x &=& 0\,, \nonumber\\
A'_y &=& d_L \tan\vth \, \cos\vphi\,, \label{eq:Acoordsy}\\
A'_z &=& d_L \tan\vth \, \sin\vphi\,,
  \label{eq:Acoords}\\
B'_x &=& 0\,, \nonumber\\
B'_y &=& d_S \tan\cb \, \cos\chi + d_{LS} \tan\vth_S \, \cos(\pi-\vphi_S)\,,
  \\
B'_z &=& d_S \tan\cb \, \sin\chi - d_{LS} \, \tan \vth_S \, \sin (\pi-\vphi_S)\,.
  \label{eq:Bcoords}
\eeqa
Substituting eqns.~(\ref{eq:Acoordsy})--(\ref{eq:Bcoords}) into eqn.~(\ref{eq:disp-def}) yields
\beqa
d_S \tan\cb \, \cos\chi & = & d_L \tan\vth \, \cos\vphi
  \ + \ d_{LS} \tan \vth_S \, \cos \vphi_S
  \ + \ \cd_y\,,
  \label{eq:le-h} \\
d_S \tan\cb \, \sin\chi & = & d_L  \tan\vth \, \sin\vphi
  \ + \ d_{LS} \tan \vth_S \, \sin \vphi_S
  \ + \ \cd_z\,.
  \label{eq:le-v}
\eeqa
The left-hand sides involve only the source position, while the
right-hand sides involve only the image position.
In other words, {\em this pair of equations
is the general form of the gravitational lens equation for
source and observer in the asymptotically flat region, for a general isolated compact object.}  Note that apart from the asymptotic flatness assumption, these equations use no properties specific to Kerr black holes; and if the deflector was a Kerr black hole, then neither the observer nor the source has been assumed to be equatorial.
We shall refer to eqns.~(\ref{eq:le-h}) and (\ref{eq:le-v}), respectively, as the
``horizonal'' and ``vertical'' components of the lens equation
due to the cosine/sine dependence on $\chi$.

Consider now the case when the light ray and
its tangent lines lie in a plane which contains the optical axis.  This forces $\chi = \vphi$ or $\chi = \vphi+\pi$ depending on whether
the source is on the same or opposite side of the lens as the image.  To distinguish these two cases, it is useful to define
$\snq = \cos(\chi-\vphi)$ to be a sign that indicates whether the
source is on the same side of the lens as the image ($\snq=+1$)
or on the opposite side ($\snq=-1$).  The condition $A' \neq B'$ will still hold in general, but the line in the lens plane from the origin to the point $B'$ will now make the same angle with respect to the $y$-axis as the point $A'$, namely, the angle $\vphi$ (see Fig.~\ref{fig:Lens-Geom3}).  As a result, the line in the source plane from the origin to the point $B''$ will also make the angle $\vphi$ with respect to the $y$-axis.  Thus we will have $\vphi_S = \vphi+\pi$.  Given these conditions,
eqns.~(\ref{eq:le-h}) and (\ref{eq:le-v}) reduce to the single lens
equation
\beq \label{eq:le-sph}
  d_S\,\snq\,\tan\cb = d_L \tan\vth - d_{LS} \tan(\hat{\alpha}-\vth) + \cd \,,
\eeq
where the displacement amplitude is $\cd = \cd_y/\!\cos\vphi = \cd_z/\!\sin\vphi$ (in the case of planar rays),
and to connect with traditional descriptions of gravitational
lensing we have introduced the light bending angle
$\hat{\alpha} \equiv \vth + \vth_S$.  (If desired, the sign $\snq$ can be
incorporated into the tangent so that the left-hand side is
written as $\tan(\snq\cb)$, where we think of $\snq\cb$ as the
signed source position.)  {\it Eqn.~(\ref{eq:le-sph}) is the general form of the lens equation in the case of planar rays.}  If the displacement $\cd$ is ignored,
then eqn.~(\ref{eq:le-sph}) matches the spherical lens equation given
by \cite{virbetal2}.  We consider the displacement term in
\refsec{spherical}.

\subsection{General Magnification Formula}
\label{sec:gen-mag}

The magnification of a small source is given by the ratio of the
solid angle subtended by the image to the solid angle subtended
by the source
(e.g.,
\cite[p.~82]{PLW}).  As measured by the observer, if $\ell$ is the
distance to the image (as opposed to the perpendicular distance),
then the small solid angle subtended by the image is
\beqa
  d\Omega_I = \frac{|(\ell \, d \vth)\, (\ell \sin\vth \, d\vphi)|}{\ell^2}
=|\sin\vth\ d\vth\ d\vphi| = |d(\cos\vth)\ d\vphi| .\nonumber
\eeqa
Similarly, the small solid angle subtended by the source is
\beqa
  d\Omega_S = |\sin\cb\ d\cb\ d\chi| = |d(\cos\cb)\ d\chi| .\nonumber
\eeqa
We then have the absolute magnification
\beqa
  |\mu| = \frac{d\Omega_I}{d\Omega_S} = |\det J|^{-1} ,\nonumber
\eeqa
where $J$ is the Jacobian matrix
\beqa
  J = \frac{\partial (\cos \cb, \chi)}{\partial(\cos\vth, \vphi)}
  = \left[\matrix{
    \frac{\partial\cos\cb}{\partial\cos\vth}
  & \frac{\partial\cos\cb}{\partial\vphi   } \cr
    \frac{\partial\chi   }{\partial\cos\vth}
  & \frac{\partial\chi   }{\partial\vphi   }
  }\right].\nonumber
\eeqa
Writing out the determinant and dropping the absolute value in
order to obtain the signed magnification, we get
\beq \label{eq:mu-general}
  \mu = \left[ \frac{\sin\cb}{\sin\vth} \left(
      \frac{\partial\cb}{\partial\vth }\ \frac{\partial\chi}{\partial\vphi}
    - \frac{\partial\cb}{\partial\vphi}\ \frac{\partial\chi}{\partial\vth }
  \right) \right]^{-1} .
\eeq
In the case of spherical symmetry, the image and source lie in
the same plane, so $\partial\cb/\partial\vphi=0$ and
$\partial\chi/\partial\vphi=1$, recovering  the familiar result
\beqa
  \mu = \left( \frac{\sin\cb}{\sin\vth}\ \frac{\partial\cb}{\partial\vth}
    \right)^{-1} .\nonumber
\eeqa

\section{Lens Equation for Kerr Black Holes}
\label{sec:lenseq-Kerr}

\subsection{Kerr Metric}
\label{sec:Kerr-metric}

\begin{figure}[t]
\begin{center}
\includegraphics[scale=.62]{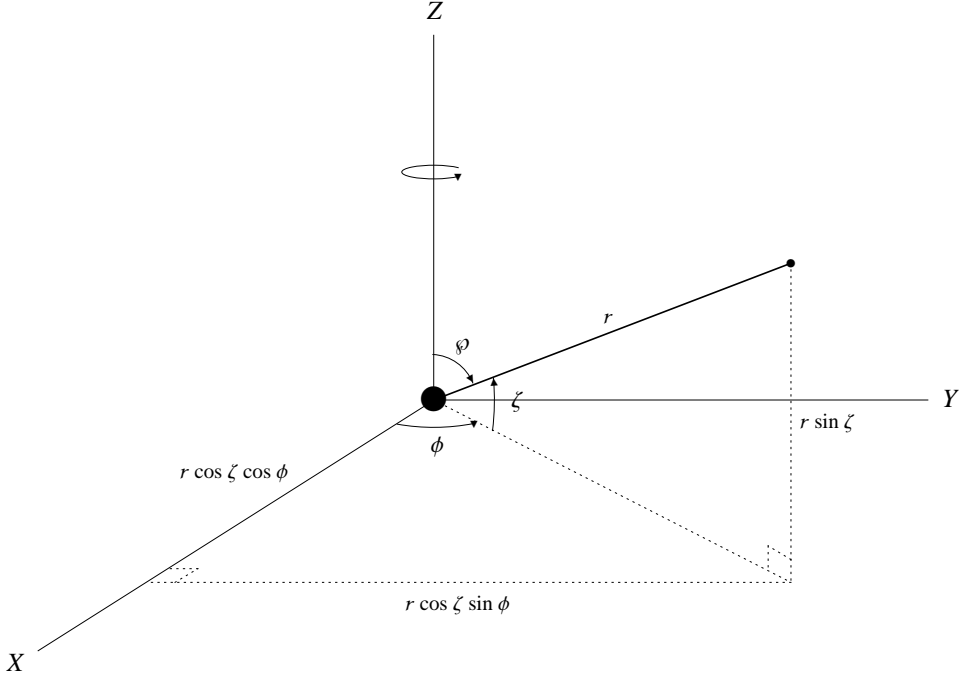}
\end{center}
\caption{
Cartesian $(X,Y,Z)$ and spherical polar $(r,\kpolar,\kazym)$
coordinates centered on the black hole, where $\kpolar = \pi/2 - \wp$ with $\wp$ the polar angle; note that $-\pi/2 \leq \kpolar \leq \pi/2$.  The black hole spins about the $Z$-axis, which corresponds to
$\kpolar=\pi/2$, in the direction of
increasing $\kazym$.  The equatorial plane
of the black hole corresponds to $\kpolar = 0$ or the
$(X,Y)$-plane.}
\label{fig:BHCoords}
\end{figure}

Now let the deflector in Fig.~\ref{fig:Lens-Geom3} be a Kerr black hole.  The Kerr metric is the unique axisymmetric,
stationary, asymptotically flat, vacuum solution of the Einstein
equations describing a stationary black hole with mass $\bhpm$
and spin angular momentum $\bhpJ$ (see, e.g., \cite[pp.~322-324]{wald}).
Consider the Kerr metric in Boyer-Lindquist coordinates
$(t,r,\wp,\kazym)$, where $\wp$ is the polar angle
and $\kazym$ the azimuthal angle.  For our purposes, it is
convenient to transform $\wp$ to $\kpolar = \pi/2-\wp$
and work with the slightly  modified Boyer-Lindquist coordinates
$(t,r,\kpolar,\kazym)$; note that $-\pi/2 \leq \kpolar \leq \pi/2$.  The spatial coordinates are shown in
\reffig{BHCoords}.

The metric takes the form
\beqa
\label{eq:app:kerr-metric}
  ds^2  = g_{tt}\, dt^2  + g_{rr}\,dr^2 
    + g_{\kpolar \kpolar}\,d\kpolar^2
    + g_{\kazym \kazym}\,d\kazym^2  
    + 2\,g_{t \kazym}\,dt\,d\kazym\,,\nonumber
\eeqa
where $t = c {\tt t}$ with ${\tt t}$ being physical time.  The
metric coefficients are
\beqa \label{eq:app:kerr-components}
  g_{tt} &=& - \frac{r(r-2\gravr) + \bha^2 \sin^2\kpolar}
    {r^2 + \bha^2 \sin^2\kpolar}\ , \label{gtt}\\
  g_{rr} &=& \frac{r^2 + \bha^2 \sin^2\kpolar}{r(r-2\gravr) + \bha^2}\ , \label{grr}\\
  g_{\kpolar \kpolar} &=& r^2 + \bha^2 \sin^2\kpolar\,, \label{gss}\\
  g_{\kazym \kazym} &=& \frac{(r^2+\bha^2)^2
    - \bha^2(\bha^2+r(r-2\gravr))\cos^2\kpolar}
    {r^2 + \bha^2 \sin^2\kpolar}\ \cos^2\kpolar\,, \label{gphiphi}\\
  g_{t \kazym} &=& - \frac{2 \gravr \bha \, r \cos^2\kpolar}
    {r^2 + \bha^2 \sin^2\kpolar}\ .\label{gtphi}
\eeqa
The parameter $\gravr$ is the gravitational radius, and $\bha$ 
is the angular momentum per unit mass:
\beqa
  \gravr = \frac{G \bhpm}{c^2}\ , \qquad
  \bha = \frac{\bhpJ}{c \bhpm}\ .\nonumber
\eeqa
Note that both $\gravr$ and $\bha$ have dimensions of length.
It is convenient to define a dimensionless spin parameter:
\beqa
\label{eq:ahat}
  \ahat = \frac{\bha}{\gravr}\ .\nonumber
\eeqa
Unless stated to the contrary, the black hole's spin is
subcritical; i.e., $\ahat^2 < 1$.

\subsection{Lens Equation for an Equatorial Observer}
\label{sec:Kerr-lens-eqn}

{\em We now specialize to the case when the observer lies
in the equatorial plane of the Kerr black hole,} so the coordinates
$(x,y,z)$ in \reffig{Lens-Geom3} coincide with the coordinates
$(X,Y,Z)$ in \reffig{BHCoords}.  Note that we still consider
general source positions.

In \refapp{Kerr-null-geodesics} we carefully analyze null
geodesics seen by an observer in the equatorial plane.  By
considering constants of the motion, we derive the following
lens equation:
\beqa
  d_S \tan\cb \cos\chi &=& d_{LS} \tan\vth_S \cos\vphi_S
    \ + \ d_L\ \frac{\sin\vth \cos\vphi}{\cos\vth_S}\ ,
    \label{eq:le-h-Kerr} \\
  d_S \tan\cb \sin\chi &=& d_{LS} \tan\vth_S \sin\vphi_S
    \ + \ \frac{d_L \sin\vth}{1 - \sin^2\vth_S \sin^2\vphi_S} \times
    \label{eq:le-v-Kerr} \\
  &&\qquad \left[ \cos\vphi \sin\vth_S \tan\vth_S \sin\vphi_S \cos\vphi_S
    + \left( \sin^2\vphi - \sin^2\vth_S \sin^2\vphi_S \right)^{1/2} \right] .
    \nonumber
\eeqa
{\it This is the general form of the lens equation for an equatorial
observer in the Kerr metric for observer and source in the
asymptotically flat region.}  
It is valid for all light rays, whether they loop
around the black hole or not, as long as they lie outside the
region of photon capture.  No small-angle approximation is required.

Note that eqns. (\ref{eq:le-h}) and (\ref{eq:le-v}) represent the general
form of the lens equation, with the displacement terms explicitly
written, while eqns. (\ref{eq:le-h-Kerr}) and (\ref{eq:le-v-Kerr}) give
the exact lens equation for an equatorial Kerr observer, with
the displacement terms implicitly included.  Demanding that these
two pairs of equations be equivalent allows us to identify the
displacement terms for an equatorial Kerr observer:
\beqa
  \cd_y &=& d_L \sin\vth \cos\vphi \left( \frac{1}{\cos\vth_S}
    - \frac{1}{\cos\vth} \right) , \label{eq:dispy}\\
  \cd_z &=& - d_L \tan\vth \sin\vphi \ + \ 
    \frac{d_L \sin\vth}{1 - \sin^2\vth_S \sin^2\vphi_S} \times \label{eq:dispz}\nonumber\\
  &&\qquad \left[ \cos\vphi \sin\vth_S \tan\vth_S \sin\vphi_S \cos\vphi_S
    + \left( \sin^2\vphi - \sin^2\vth_S \sin^2\vphi_S \right)^{1/2} \right].\label{eq:dispz}
\eeqa

\subsection{Schwarzschild Case}
\label{sec:spherical}

In the case of a spherically symmetric lens we have $\vphi_S = \vphi + \pi$, and either $\chi = \vphi$
or $\chi = \vphi+\pi$, depending on whether the source lies on the
same or opposite side of the lens as the image.  Once again, we define $\snq = \cos(\chi-\vphi)$ to be
a sign that distinguishes these two cases.  With these conditions
eqns.~(\ref{eq:le-h-Kerr}) and (\ref{eq:le-v-Kerr}) combine to form the
single lens equation with displacement for a Schwarzschild black hole:
\beqa \label{eq:le-sph-Kerr}
  d_S\,\snq\,\tan\cb = d_L\ \frac{\sin\vth}{\cos\vth_S} \ - \ 
    d_{LS}\,\tan\vth_S\,.
\eeqa
Two comments are in order.  First, our spherical lens equation
(\ref{eq:le-sph-Kerr}) is equivalent to the spherical lens
equation recently derived by Bozza \& Sereno \cite{BS,Bozza2} (up to the sign $\snq$,
which was not discussed explicitly in \cite{BS,Bozza2}).
The second comment refers to the amplitude of the displacement.
By comparing our general planar-ray lens equation (\ref{eq:le-sph}) with eqn.~(\ref{eq:le-sph-Kerr}),
we can identify the displacement
\beqa
\label{eq:disp-sph}
  \cd = d_L \sin\vth \left[ \frac{1}{\cos(\alpha-\vth)} - \frac{1}{\cos\vth}
    \right]\,,
\eeqa
where we have switched from $\vth_S$ to the bending angle
$\alpha = \vth+\vth_S$.  Now let
$\delta\alpha = \alpha \, {\rm mod} \, 2 \pi$, and assume that
$\vth$ and $\delta\alpha$ are small.  (Note that we need not
assume $\alpha$ itself is small, only that $\delta\alpha$ is small.
This means that our analysis applies to all light rays, including those that loop
around the lens.)  Taylor expanding the displacement in $\vth$
and $\alpha$ yields
\beqa
  \cd = \frac{d_L}{2} \, (\vth \ \delta \alpha) 
         (\delta \alpha - 2 \vth)  \ + \ {\cal O}(4).\nonumber
\eeqa

\section{Conclusions}
\label{sec:conclusions}

Recently a lens equation was developed for Schwarzschild lensing with displacement (see Section~\ref{sec:spherical} above), when the light ray's tangent lines
at the source and observer do not meet on the lens plane.  In this paper we found a new generalization of the lens equation with displacement for axisymmetric lenses that extends the previous work to a fully three-dimensional setting.  Our formalism assumes that the source and observer are in the asymptotically flat region, and does not require a small angle approximation.  Furthermore, we found a new magnification formula applicable to this more general context.  Our lens equation is thus applicable to non-spherically symmetric compact bodies, such as Kerr black holes.  We gave explicit formulas for the
displacement when the observer is in the equatorial plane
of a Kerr black hole and 
for the situation of spherical symmetry.

\begin{acknowledgments}

ABA and AOP would especially like to thank Marcus C. Werner for helpful discussions.  AOP acknowledges the support of NSF Grant DMS-0707003.

\end{acknowledgments}

\appendix


\section{Exact Kerr Null Geodesics for Equatorial Observers}
\label{app:Kerr-null-geodesics}

In this Appendix we determine the equations of motion governing
light rays seen by equatorial observers in the Kerr metric.  We
obtain the exact equations of motion by considering constants of
the motion.

\subsection{Equations of Motion for Null Geodesics}


We first study the equations of motion for a general Kerr null geodesic
$\cur(\lambda) = (t(\lambda),r(\lambda),\kpolar(\lambda),\kazym(\lambda))$,
where $\lambda$ is an affine parameter.  The geodesic is assumed
to be outside the region of photon capture.  Two
immediate constants of the motion for Kerr geodesics are the energy
$\cE$ and the orbital angular momentum $\cL$.  They yield two equations
of motion (see, e.g., \cite[p. 180]{Oneill}):  
\beqa
  \dot{t} &=& \frac{g_{\kazym \kazym} \cE + g_{t \kazym} \cL}
    {g_{t \kazym}^2 - g_{t t} g_{\kazym \kazym}}\ , \label{eq:app:tdot1} \nonumber\\
  \dot{\kazym} &=& \frac{g_{t \kazym} \cE + g_{t t} \cL}
    {g_{t t} g_{\kazym \kazym} - g_{t \kazym}^2}\ , \label{eq:app:xdot1}\nonumber
\eeqa
where the dot denotes differentiation relative to the affine
parameter $\lambda$.  Since we only consider unbound light rays,
we may assume $\cE > 0$.  With a suitable fixed choice of affine
parameter $\lambda$, we henceforth assume that $\cE\, \lambda$
has dimension of length.  The dimension of the ratio $\cL/\cE$
is also length.  A third constant of motion is nullity, which yields
\beqa
\label{eq:app:rdot1}
  \dot{r} = \pm \left( \frac{
    -g_{t    t   } \dot{t}^2
    -g_{\kpolar  \kpolar } \dot{\kpolar}^2
    -g_{\kazym \kazym} \dot{\kazym}^2
    -2 g_{t \kazym} \dot{t} \dot{\kazym}}{g_{rr}} \right)^{1/2} .\nonumber
\eeqa
A fourth constant of motion $\cQ$ is the Carter constant, which
comes from separating the Hamilton-Jacobi equation \cite{carter}.   We henceforth assume that $\cQ \geq 0$; i.e., that the light ray either crosses the equatorial plane or asymptotically approaches it (see, e.g., \cite[pp.~204-205]{Oneill}).   Employing the notation
\beqa
  \hl = \cE \, \lambda\,, \qquad 
  \hat{\dot{x}} = \frac{d x}{d \hl}\ , \qquad
  \hcL = \frac{\cL}{\cE}\ , \qquad
  \hcQ = \frac{\cQ}{\cE^2}\ ,\nonumber
\eeqa
the fourth equation of motion can be written
\beqa
  \hat{\dot{\kpolar}} = \frac{d \kpolar}{d \hl} = \pm \frac{( \hcQ + \bha^2 \sin^2\kpolar
    - \hcL^2 \tan^2\kpolar )^{1/2}}{r^2 + \bha^2 \sin^2\kpolar}\ .
    \label{eq:app:edot}
\eeqa
Using the metric coefficients (\ref{gtt})--(\ref{gtphi}) shown in Section~\ref{sec:Kerr-metric}, the null geodesic
equations of motion become
\beqa
  \hat{\dot{t}} &=& 1 + \frac{2\gravr\,r(\bha^2-\bha\hcL+r^2)}
    {[\bha^2+r(r-2\gravr)] (r^2 + \bha^2 \sin^2\kpolar)}\ ,
    \label{eq:app:tdot} \\
  \hat{\dot{r}} &=& \pm \frac{[ r^4 - (\hcQ + \hcL^2-\bha^2) r^2
    + 2\gravr ((\hcL - \bha)^2+ \hcQ) r - \bha^2 \hcQ ]^{1/2}}
    {r^2 + \bha^2 \sin^2\kpolar}\ ,
    \label{eq:app:rdot} \\
  \hat{\dot{\kazym}} &=& \frac{ 2 \bha \gravr r
    + \hcL r (r-2\gravr) \sec^2\kpolar + \bha^2 \hcL \tan^2\kpolar }
    {[\bha^2+r(r-2\gravr)] (r^2 + \bha^2 \sin^2\kpolar)}\ .
    \label{eq:app:xdot}
\eeqa
Eqns.~(\ref{eq:app:edot})--(\ref{eq:app:xdot}) form the set
of equations of motion that we must solve in order to describe null
geodesics in the Kerr metric.

\subsection{Exact Lens Equation (with Displacement) for Equatorial Observers}
Assuming that the source and equatorial observer are in the asymptotically
flat region, we 
consider now the constants of motion $\hcL$ and $\hcQ$.  We
can find them by examining the equations of motion in the
asymptotically flat region of the spacetime far from the black
hole.  Formally, this means taking the limits $\bha, \gravr \to 0$, in which case the equations of motion reduce to
\beq \label{eq:app:EofMflat}
  \hat{\dot{t}} = 1\ , \quad
  \hat{\dot{r}} = \pm \frac{(r^2 - \hcQ - \hcL^2)^{1/2}}{r}\ , \quad
  \hat{\dot{\kpolar}} = \pm \frac{(\hcQ - \hcL^2\tan^2\kpolar)^{1/2}}{r^2}\ ,
  \quad
  \hat{\dot{\kazym}} = \frac{\hcL}{r^2 \cos^2\kpolar}\ .
\eeq
At the position of the observer, the light ray is a straight
line described by the angles $\vth$ and $\vphi$.  For an
equatorial observer, we see from \reffig{Lens-Geom3} that the three Cartesian components of the
line can be written as
\beqa
  x(\hl) &=& d_L + (\hl -\hl_O) \cos\vth\,,      \nonumber\\
  y(\hl) &=& -(\hl - \hl_O) \sin\vth \cos\vphi\,, \nonumber\\
  z(\hl) &=& -(\hl - \hl_O) \sin\vth \sin\vphi\,,\nonumber
\eeqa
where $\hl_O$ is the value of the affine parameter at the
position of the observer, and the affine parameter range for the line segment is $-d_L/\!\cos\vth + \hl_O \leq \hl \leq \hl_O$ (recall that $0 < \vth < \pi/2$ and that $\hl$ has dimension of length).  Next, we convert to spherical coordinates
$(r, \kpolar, \kazym)$ and evaluate $r$, $\kpolar$, $\hat{\dot{r}}$, and $\hat{\dot{\kazym}}$ at $\hl = \hl_O$:
\beqa
\label{hl_O}
r(\hl_O) = d_L\ ,\quad \kpolar(\hl_O) = 0\ ,\quad \hat{\dot{r}}\,(\hl_O) = \,\cos\vth\ ,\quad \hat{\dot{\kazym}}\,(\hl_O) = \,-\frac{\sin\vth\cos\vphi}{d_L}\ . 
\eeqa
Finally, we substitute eqn.~(\ref{hl_O}) into eqn.~(\ref{eq:app:EofMflat})
to solve for $\hcL$ and $\hcQ$ when $\hl = \hl_O$:
\beqa
  \hcL_O &=& \hat{\dot{\kazym}}\,r^2\cos^2\kpolar\,\Big|_{\hl = \hl_O} = - d_L \sin\vth \cos\vphi\,,\label{eq:LQ-O1}\\
  \hcQ_O &=& \left[r^2\left(1 - \hat{\dot{r}}^{\,2}\right) - \hcL^2\right]\Big|_{\hl = \hl_O} = d_L^2 \sin^2\vth \sin^2\vphi\,.\label{eq:LQ-O2}
\eeqa
(To be clear, we have labeled these constants of motion with ``O''
for observer.  Note that $\hcQ_O \geq 0$.  Note also that we could just as well have used $\hat{\dot{r}}$ and $\hat{\dot{\kpolar}}$, or $\hat{\dot{\kpolar}}$ and $\hat{\dot{\phi}}$, to solve for $\hcL$ and $\hcQ$.)  Going further, we define
\beqa
\label{eq:impact}
  b \equiv \left(\hcQ + \hcL^2\right)^{1/2} = d_L \sin\vth
\eeqa
to be the (absolute) impact parameter of the light ray (since $0 < \vth < \pi/2$, $\sin\vth$ is positive.)
This is clearly a constant of the motion.

We could equally well express the constants of motion in terms
of the light ray at the position of the source.  To be clear,
we write these constants as $\hcL_S$ and $\hcQ_S$.  Recall that the position
of the source is defined by the angles $(\cb,\chi)$, while the
direction of the light ray at the source is defined by the angles
$(\vth_S,\vphi_S)$.  So the three Cartesian components of the
light ray at the source can be written as
\beqa
  x(\hl) &=& -d_{LS} + (\hl - \hl_S) \cos\vth_S\,, \nonumber\\
  y(\hl) &=& d_S\,Y_{\rm{src}}  - (\hl - \hl_S) \sin\vth_S \cos\vphi_S\,,
    \nonumber\\
  z(\hl) &=& d_S\,Z_{\rm{src}}  - (\hl - \hl_S) \sin\vth_S \sin\vphi_S\,,\nonumber
\eeqa
where $\hl_S$ is the value of the affine parameter at the
position of the source, and the affine parameter range for the line segment is $\hl_S \leq \hl \leq d_{LS}/\!\cos\vth_S + \hl_S$ (recall that $0 \leq \vth_S < \pi/2$).  Note that for simplicity we have defined
\beqa 
\label{eq:app:YZ}
  Y_{\rm{src}} \equiv \tan\cb \cos\chi\,, \qquad
  Z_{\rm{src}} \equiv \tan\cb \sin\chi\,.\nonumber
\eeqa
By a computation identical to those in eqns.~(\ref{hl_O}) and (\ref{eq:LQ-O2}), we solve eqn.~(\ref{eq:app:EofMflat}) for $\hcL$ and $\hcQ$ when $\hl = \hl_S$ to obtain
\beqa
  \hcL_S &=& - d_S\,Y_{\rm{src}} \cos\vth_S + d_{LS} \sin\vth_S \cos\vphi_S\,,
     \label{eq:LQ-S1}\\
  \hcQ_S &=& d_S^2 Z_{\rm{src}}^2 ( \cos^2\vth_S + \sin^2\vth_S \cos^2\vphi_S )
  \nonumber\\
  &&
    - 2\,d_S\,Z_{\rm{src}}\,\sin\vth_S \sin\vphi_S ( d_{LS} \cos\vth_S
      + d_S\,Y_{\rm{src}} \sin\vth_S \cos\vphi_S )
  \nonumber\\
  &&
    + (d_S^2 Y_{\rm{src}}^2 + d_{LS}^2) \sin^2\vth_S \sin^2\vphi_S\,.
    \label{eq:LQ-S2}
\eeqa
Since we are discussing constants of the motion, we must have
$\hcL_O = \hcL_S$ and $\hcQ_O = \hcQ_S$.  Using eqns.~(\ref{eq:LQ-O1}) and (\ref{eq:LQ-S1}), the condition $\hcL_O = \hcL_S$ is a trivial
linear equation for $Y_{\rm{src}}$, which yields
\beqa
d_S \tan\cb \cos\chi = d_{LS} \tan\vth_S \cos\vphi_S
    \ + \ d_L\ \frac{\sin\vth \cos\vphi}{\cos\vth_S}\ . \label{lens1}
\eeqa
Using eqns.~(\ref{eq:LQ-O2}) and (\ref{eq:LQ-S2}), the condition $\hcQ_O = \hcQ_S$ is
a quadratic equation in $Z_{\rm{src}}$, which yields the following two roots:
\beqa
  Z_{\rm{src}} &=& \frac{d_{LS} \tan\vth_S \sin\vphi_S}{d_S}
    \ + \ \frac{d_L \sin\vth}{d_S\left(1 - \sin^2\vth_S \sin^2\vphi_S\right)} \times \label{lens2}\\
  &&\qquad \left[ \cos\vphi \sin\vth_S \tan\vth_S \sin\vphi_S \cos\vphi_S
    \pm \left( \sin^2\vphi - \sin^2\vth_S \sin^2\vphi_S \right)^{1/2} \right] .
    \nonumber
\eeqa
%
%
%
We will take the positive root in eqn.~(\ref{lens2}) because in the case of spherical symmetry only the positive root, taken together with eqn.~(\ref{lens1}), will combine to form eqn.~(\ref{eq:le-sph-Kerr}) in Section~\ref{sec:spherical}.  We thus have
\beqa
  d_S\tan\cb \sin\chi &=& d_{LS} \tan\vth_S \sin\vphi_S
    \ + \ \frac{d_L \sin\vth}{1 - \sin^2\vth_S \sin^2\vphi_S} \times \label{lens3}\\
  &&\qquad \left[ \cos\vphi \sin\vth_S \tan\vth_S \sin\vphi_S \cos\vphi_S
    + \left( \sin^2\vphi - \sin^2\vth_S \sin^2\vphi_S \right)^{1/2} \right] .
    \nonumber
\eeqa
{\it Eqns.~(\ref{lens1}) and (\ref{lens3}) thus constitute the two components of the general lens
equation for an equatorial observer in the Kerr metric.}

%
%
%
%

{}

\end{document}